\documentstyle[epsfig,feynmp]{aipproc}

\newcommand{\beq}{\begin{eqnarray}}
\newcommand{\eeq}{\end{eqnarray}}
\newcommand{\str}{\vphantom{\bigg(}}
\newcommand{\s}{\\ \vspace*{-3.5mm} }

\begin{document}
\hspace{0.9cm}\rightline{
        \begin{minipage}{4cm}
        PM/01--03\\
        hep-ph/0101262\hfill \\
        \end{minipage}        
}

\title{Testing Higgs Self-Couplings at High-Energy Linear Colliders}
\author{M.M.~M\"uhlleitner}
\address{Laboratoire de Physique Math\'ematique et Th\'eorique, 
UMR5825--CNRS, Universit\'e de Montpellier II, F--34095 Montpellier Cedex 5,
France}
\maketitle
\vspace{-0.5cm}

\begin{abstract}
In order to verify the Higgs mechanism experimentally, the Higgs self-couplings have to be probed. These couplings allow the reconstruction of the characteristic Higgs potential responsible for the electroweak symmetry breaking. The couplings are accessible in a variety of multiple Higgs production processes. The theoretical analysis including the most relevant channels for the production of neutral Higgs boson pairs at high-energy and high-luminosity $e^+e^-$ linear colliders will be presented in this note.
\end{abstract}
\vspace{-0.2cm}

{\bf 1.} Within the Higgs mechanism \cite{higgs} the electroweak gauge bosons 
and fundamental matter particles acquire their masses through the interaction 
with a scalar field. The self-interaction of the scalar field induces, via a 
non-vanishing field strength $v=(\sqrt{2}G_F)^{-1/2} \approx 246$~GeV, the 
spontaneous breaking of the electroweak $SU(2)_L\times U(1)_Y$ symmetry down 
to the electromagnetic $U(1)_{EM}$ symmetry. 

To establish the Higgs mechanism experimentally, the self-energy potential of the Standard Model \cite{gunion}, $V=\lambda \left( \Phi^\dagger \Phi - v^2/2 \right)^2$, with a minimum at $\langle\Phi \rangle_0=v/\sqrt{2}$ must be reconstructed. This task requires the measurement of the Higgs self-couplings of the physical Higgs boson $H$, which can be read off directly from the potential
\beq
V = \frac{M_H^2}{2} H^2 + \frac{M_H^2}{2v} H^3 
+ \frac{M_H^2}{8v^2} H^4
\label{hpot}
\eeq
As evident from Eq.~(\ref{hpot}), in the SM the trilinear and quadrilinear 
vertices are uniquely determined by the mass of the Higgs boson, 
$M_H = \sqrt{2\lambda}v$. 

\begin{fmffile}{fd}
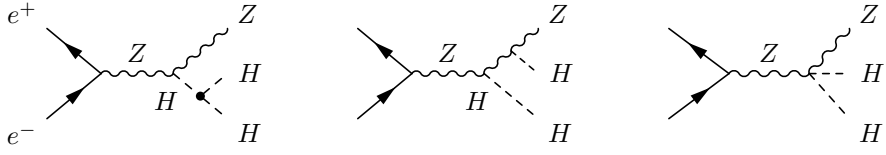
\begin{figure}
\begin{flushleft}
{\footnotesize
\unitlength1mm
\hspace{10mm}
\begin{center}
\begin{fmfshrink}{0.7}
\begin{fmfgraph*}(24,12)
  \fmfstraight
  \fmfleftn{i}{3} \fmfrightn{o}{3}
  \fmf{fermion}{i1,v1,i3}
  \fmflabel{$e^-$}{i1} \fmflabel{$e^+$}{i3}
  \fmf{boson,lab=$Z$,lab.s=left,tens=3/2}{v1,v2}
  \fmf{boson}{v2,o3} \fmflabel{$Z$}{o3}
  \fmf{phantom}{v2,o1}
  \fmffreeze
  \fmf{dashes,lab=$H$,lab.s=right}{v2,v3} \fmf{dashes}{v3,o1}
  \fmffreeze
  \fmf{dashes}{v3,o2} 
  \fmflabel{$H$}{o2} \fmflabel{$H$}{o1}
  \fmfdot{v3}
\end{fmfgraph*}
\hspace{15mm}
\begin{fmfgraph*}(24,12)
  \fmfstraight
  \fmfleftn{i}{3} \fmfrightn{o}{3}
  \fmf{fermion}{i1,v1,i3}
  \fmf{boson,lab=$Z$,lab.s=left,tens=3/2}{v1,v2}
  \fmf{dashes}{v2,o1} \fmflabel{$H$}{o1}
  \fmf{phantom}{v2,o3}
  \fmffreeze
  \fmf{boson}{v2,v3,o3} \fmflabel{$Z$}{o3}
  \fmffreeze
  \fmf{dashes}{v3,o2} 
  \fmflabel{$H$}{o2} \fmflabel{$H$}{o1}
\end{fmfgraph*}
\hspace{15mm}
\begin{fmfgraph*}(24,12)
  \fmfstraight
  \fmfleftn{i}{3} \fmfrightn{o}{3}
  \fmf{fermion}{i1,v1,i3}
  \fmf{boson,lab=$Z$,lab.s=left,tens=3/2}{v1,v2}
  \fmf{dashes}{v2,o1} \fmflabel{$H$}{o1}
  \fmf{dashes}{v2,o2} \fmflabel{$H$}{o2}
  \fmf{boson}{v2,o3} \fmflabel{$Z$}{o3}
\end{fmfgraph*}
\end{fmfshrink}
\end{center}
}
\end{flushleft}
\vspace{0.7cm}
\caption{Subprocesses contributing to double Higgs-strahlung, $e^+e^-\to ZHH$, in the Standard Model at $e^+e^-$ linear colliders.}
\end{figure}
\end{fmffile}
The trilinear self-coupling $\lambda = 6\sqrt{2}\lambda$ in units of $v/\sqrt{2}$ is accessible directly in Higgs pair production at high-energy $e^+e^-$ linear colliders. For c.m.\ energies up to about 1~TeV, double Higgs-strahlung \cite{hrad1,hrad2}
\beq
e^+e^- \to ZHH
\eeq
is the most promising process \cite{muehlmm}. The process includes the amplitude involving the trilinear Higgs self-coupling and two additional amplitudes due to the standard 
\begin{figure}[t]
\begin{center}
\epsfig{figure=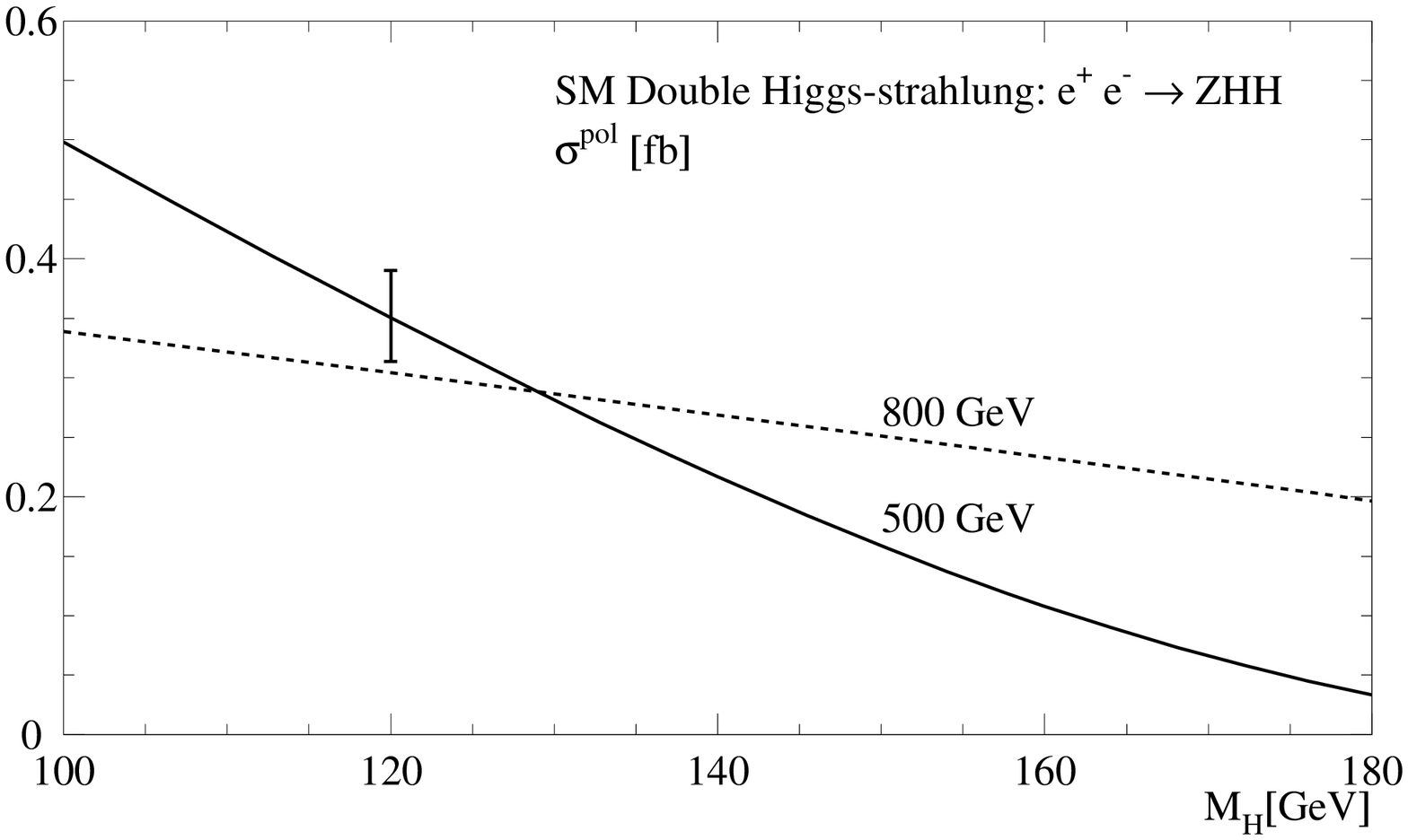,width=6.8cm}
\hspace{8mm}
\epsfig{figure=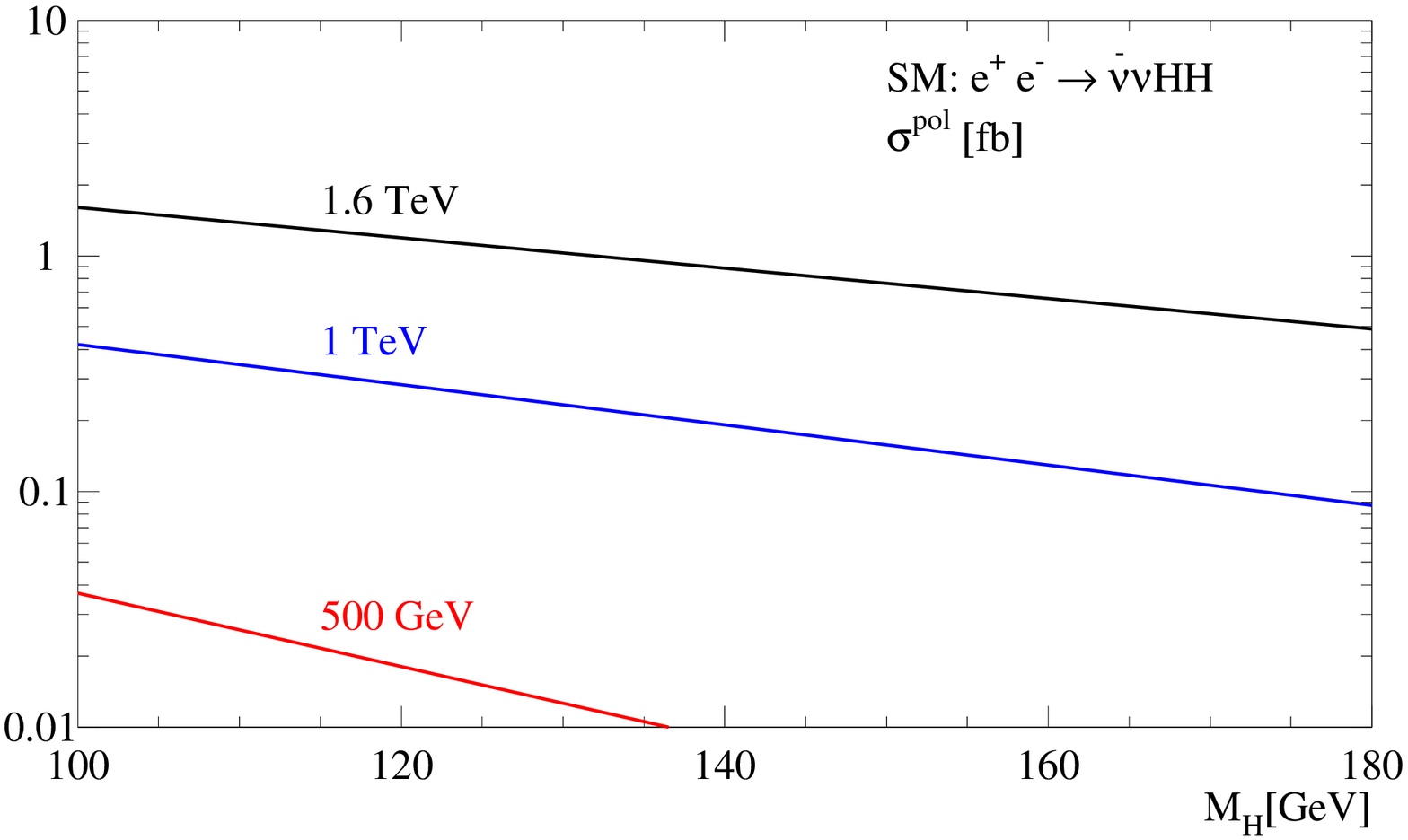,width=6.8cm}
\end{center}
\caption[]{(a) The cross section for double Higgs-strahlung, $e^+e^-\to ZHH$, in the Standard Model at two collider energies: $\sqrt{s}=500$~GeV and $800$~GeV. The electron/positron beams are taken oppositely polarized. The vertical bar corresponds to a variation of the trilinear Higgs coupling between $0.8$ and $1.2$ of the SM value.  (b) The cross section for $WW$ double Higgs fusion, $e^+e^-\to \nu_e\bar{\nu}_e HH$, at $\sqrt{s}=500$~GeV, 1~TeV and 1.6~TeV. The initial beams are polarized.}
\end{figure}
electroweak gauge interactions, cf.~Fig.~1, so that it is a binomial in $\lambda_{HHH}$. As evident from Fig.~2a the cross section is very sensitive to the trilinear Higgs self-coupling and amounts up to 0.35~fb for $M_H=120$~GeV and a c.m.~energy of 500~GeV. Scaling with the energy, it decreases to $0.3$~fb at $\sqrt{s}=800$~GeV. Experimental detector simulations of signal and background processes in the SM have demonstrated that the Higgs self-coupling can be extracted with an accuracy of $\sim 20\%$ for $M_H=120$~GeV at high luminosity $\int{\cal L} = 2$~ab$^{-1}$ \cite{gay}.

The $WW$ double Higgs fusion process \cite{hrad2,wwfus}
\beq
e^+e^-\to \nu_e\bar{\nu}_e HH
\eeq
which increases with rising $\sqrt{s}$, can be exploited for larger energies, cf.~Fig.~2b. [$\sigma = 0.37$~fb for $M_H=120$~GeV and $\sqrt{s}=1$~TeV, polarized $e^+e^-$ beams.]

Triple Higgs production is sensitive to the quadrilinear Higgs self-coupling. Due to the suppressed coupling and an additional particle in the final state, the cross section $\sigma(e^+e^-\to ZHHH)$ is only of $\cal{O}($ab$)$ and therefore not measurable at typical linear collider energies and luminosities \cite{muehlmm}. \s
%

{\bf 2.} In the Minimal Supersymmetric Extension of the Standard Model (MSSM) with five physical Higgs particles $h,H,A$ and $H^\pm$ \cite{habergun}, a plethora of trilinear and quadrilinear Higgs self-couplings can be realized. The CP-invariant couplings 
\begin{table}
\begin{center}$
\renewcommand{\arraystretch}{1.3}
\begin{array}{|l||cccc|c||ccc|}\hline
\phantom{\lambda} & 
\multicolumn{4}{|c|}{\mathrm{double\;Higgs\!-\!strahlung}} &
\multicolumn{4}{|c|}{\phantom{d} \mathrm{triple\;Higgs\!-\!production 
\phantom{d}}} \str \\
\phantom{\lambda i}\lambda & Zhh & ZHh & ZHH & ZAA & 
\multicolumn{2}{|c}{\phantom{d}Ahh \phantom{d}AHh} & \phantom{d}AHH & \!\! AAA
 \\ \hline\hline
hhh & \times & & & & \phantom{d}\times\phantom{d} & & &  \\
Hhh & \times & \times & & & \times & \times & & \\
HHh & & \times & \times & & & \times & \times & \\ 
HHH & & & \times & & & & \times & \\ 
\cline{1-6} & & & & & \multicolumn{2}{c}{\phantom{\times}} & & \\[-0.575cm]
\hline
hAA & & & & \times & \multicolumn{2}{c}{\,\,\,\times \quad\quad\, \times} & & 
\times \\ 
HAA & & & & \times & \multicolumn{2}{c}{\phantom{\times}\quad\quad\,\,\,\, 
\times}
 & \times & \times \\
\hline
\end{array}$
\end{center}
\renewcommand{\arraystretch}{1}
\vspace{-0.5cm}
\end{table}
{\footnotesize {\bf TABLE 1}: The trilinear couplings between neutral CP-even and CP-odd MSSM Higgs bosons, which can generically be probed in double Higgs-strahlung and associated triple Higgs-production, are marked by a cross. [The matrix for WW fusion 
is isomorphic to the matrix for Higgs-strahlung.]}
\smallskip\\[0.3cm] 
among the neutral Higgs bosons, $\lambda_{hhh}$, $\lambda_{Hhh}, 
\lambda_{HHh}, \lambda_{HHH}, \lambda_{hAA}, \lambda_{HAA}$, are involved 
in a large number of processes \cite{muehlmm,mssmproc}. The double and triple Higgs production processes and the trilinear couplings, that can be probed in the respective process, are listed in Table~1. The system is solvable for all $\lambda$'s up to discrete ambiguities. However, in practice, not all processes are accessible experimentally so that one has to follow the reverse direction in this case: Comparing the theoretical predictions with the experimental results of the accessible channels, the trilinear Higgs self-couplings can be tested stringently. 

The process $e^+e^- \to Zhh$ is sensitive to the trilinear coupling of the
 light CP-even Higgs boson $h$,
\beq
\lambda_{hhh} = 3\cos2\alpha\sin(\beta+\alpha) + {\cal O}(G_F M_t^4/M_Z^2)
\eeq
expressed in the mixing angles $\alpha$ and $\beta$, in a large range of the 
MSSM parameter space, as can be inferred from Fig.~3. It shows the 
$2\sigma$ sensitivity area in the $[M_A,\tan\beta]$ plane for a non-zero 
coupling at an integrated luminosity of 2 ab$^{-1}$. The cross section is 
required to exceed 0.01~fb. The sensitivity areas are significantly smaller 
for processes involving heavy Higgs bosons $H$ and $A$ in the final state. 
Details can be found in Ref.\cite{muehlmm}. \s
\begin{figure}[t]
\begin{center}
\epsfig{figure=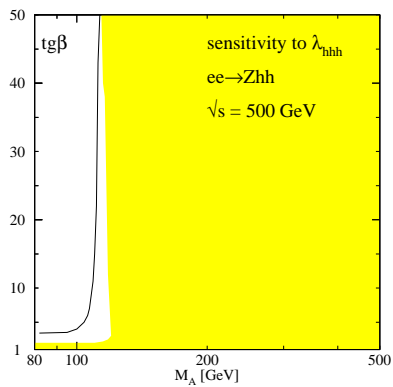,width=5.1cm}
\end{center}
\caption[]{MSSM: Sensitivity to the coupling $\lambda_{hhh}$ of the light CP-even neutral Higgs boson $h$ in the process $e^+e^-\to Zhh$ for a collider energy of $500$~GeV (no mixing).}
\end{figure}

{\bf 3.} {\bf In summary.} The measurement of the Higgs self-couplings is essential for the reconstruction of the characteristic self-energy potential. The large luminosities, which are available at future high-energy $e^+e^-$ linear colliders, allow the measurement of the trilinear Higgs self-couplings via double Higgs-strahlung and $WW$ double Higgs fusion.\s

{\bf Acknowledgements.} I am grateful to my collaborators A.~Djouadi, W.~Kilian and P.M.~Zerwas for helpful discussions. Further more I would like to thank the organizers of the Linear Collider Workshop at Fermilab Oct 2000 for the nice atmosphere. This work is supported by the European Union under contract HPRN-CT-2000-00149.

\end{document}